\documentclass[%
 reprint,
 amsmath,amssymb,
 aps,
]{revtex4-2}
\usepackage[english]{babel}
\usepackage{graphicx}
\usepackage{mathalpha}
\usepackage{cancel}
\usepackage{bbm}
\usepackage{tabularx}
\usepackage{tikz}
\usepackage{tipa}
\usetikzlibrary{matrix, backgrounds}
\usepackage{amsthm}
\usetikzlibrary{external}
\definecolor{lmugreen}{RGB}{0, 136, 58}
\definecolor{lmublack}{RGB}{35,35,35}

\definecolor{lmudarkgrey}{RGB}{98, 100, 104}
\definecolor{lmudarkgray}{RGB}{98, 100, 104}

\definecolor{lmumiddlegrey}{RGB}{192, 193, 195}
\definecolor{lmumiddlegray}{RGB}{192, 193, 195}

\definecolor{lmulightgrey}{RGB}{230, 230, 231}
\definecolor{lmulightgray}{RGB}{230, 230, 231}

\definecolor{lmubrightgrey}{RGB}{245, 245, 245}
\definecolor{lmubrightgray}{RGB}{245, 245, 245}

\definecolor{lmuaccentblue}{RGB}{15, 25, 135}
\definecolor{lmuaccentcyan}{RGB}{100, 59, 227}
\definecolor{lmuaccentviolet}{RGB}{140, 64, 145}
\definecolor{lmuaccentred}{RGB}{215, 25, 25}
\definecolor{lmuaccentorange}{RGB}{241, 135, 0}

\pgfdeclarelayer{bottom}
\pgfsetlayers{bottom, background, main}

\tikzset{gate/.style={rectangle, draw=black, fill=white, text height=1.4ex, yshift=-2pt}, every matrix of nodes/.style={matrix of nodes, nodes in empty cells}, every matrix/.style={nodes in empty cells, row sep=.2cm, column sep=.1cm}}

\newcommand{\SWAP}{\operatorname{SWAP}}
\newcommand{\CNOT}{\operatorname{CNOT}}

\theoremstyle{definition}
\newtheorem{definition}{Definition}[section]
\newtheorem{proof2}{Proof}[section]

\begin{document}

\title{Permutation-invariant quantum circuits}

\author{Maximilian Balthasar Mansky}%
 \email{maximilian-balthasar.mansky@ifi.lmu.de}
  \affiliation{Department of Informatics, LMU Munich, 80538 Munich, Germany}
 \author{Santiago Londoño Castillo}
  \affiliation{Department of Informatics, LMU Munich, 80538 Munich, Germany}
 \author{Victor Ramos Puigvert}
  \affiliation{Department of Informatics, LMU Munich, 80538 Munich, Germany}
\author{Claudia Linnhoff-Popien}
 \affiliation{Department of Informatics, LMU Munich, 80538 Munich, Germany}

\date{\today}

\begin{abstract}
The implementation of physical symmetries into problem descriptions allows for the reduction of parameters and computational complexity. We show the integration of the permutation symmetry as the most restrictive discrete symmetry into quantum circuits. The permutation symmetry is the supergroup of all other discrete groups. We identify the permutation with a $\SWAP$ operation on the qubits. Based on the extension of the symmetry into the corresponding Lie algebra, quantum circuit element construction is shown via exponentiation. This allows for ready integration of the permutation group symmetry into quantum circuit ansatzes. The scaling of the number of parameters is found to be $\mathcal{O}(n^3)$, significantly lower than the general case and an indication that symmetry restricts the applicability of quantum computing. We also show how to adapt existing circuits to be invariant under a permutation symmetry by modification.
\end{abstract}

\maketitle


\section{Introduction}
The mathematical study of group symmetries has proven to be one of the most successful tools in mathematical physics. From the development of quantum chromodynamics \cite{Gell-Mann:1961omu} from $\operatorname{SU}(3)$ symmetries to the discovery of exact solutions to Einstein's equations from their respective isometry groups \cite{Stephani:2003tm} and advancements in string theory from duality symmetries \cite{delaOssa:1992vci}, there exist many instances in physics where the study of the symmetries of a problem led to its resolution or even to the development of an entirely new theory.

Group theory has developed out of the consideration of physical symmetries and provides an accurate description of the behaviour of the system under symmetry operations \cite{sternberg_group_1995}. In particular, it gives a notion of systems that do not change under a symmetry operation and thus are symmetry-invariant. Computational problems from other fields of science can similarly utilize the description of symmetry operations to simplify the problem and reduce the solution space \cite{hamermesh_group_2012}.

The consideration of symmetries is in general independent of the computational method. For classical computers, an appropriate design of the computational algorithm allows to capture the underlying symmetry of the solution to be found. On quantum computers, the discrete symmetry can be integrated into the quantum circuit \cite{larocca_group-invariant_2022, meyer_exploiting_2023}. Quantum computers are structurally beneficial for the study of quantum systems \cite{georgescu_quantum_2014} and as such need to be able to capture symmetries within the system of study.

For a study of a physical system, it is often helpful to look at the extremes first. Quantum circuit construction in the absence of external symmetries is possible, both towards a known target \cite{mansky_near-optimal_2023} and also with a quantum machine learning ansatz \cite{schuld_introduction_2015}. On the other end, the permutation symmetry acts as the supergroup of all other discrete symmetries. In the quantum computing case, this can be realized with the requirement that all qubits are interchangeable with all other qubits. This assumes that the smallest unit of information processed in the system is a single qubit. It is also possible to require larger blocks of qubits representing a block of information to be interchangeable, or consider the interchangeability of logical qubits \cite{nigg_quantum_2014}. 

By embedding the physical problem in the description of the quantum computer, one also imparts additional knowledge on the system. There is always a trade-off between generalization and specialization of systems \cite{giles_learning_1987}. This is also true for implementing symmetries on a computational problem, as the solution is then inherently limited to that symmetry-respecting subspace of the total solution space. In the case where the implemented symmetry does not correspond to the factual symmetry of the problem, the solution can only be approximative. On the other hand, if the factual symmetry is correctly implemented, the lower number of parameters should lead to a faster convergence. In the context of quantum computing, the symmetry restricts the size of the available Hilbert space \cite{ragone_unified_2023}. 

The structure of the paper is as follows. After an overview of prior and related work, we first introduce the mathematical fundamentals and show that a permutation-invariant circuit is well-defined, in terms of its group structure and corresponding Lie algebra, and has the necessary structure for quantum computing. A constructive approach for permutation invariant circuits is introduced in section \ref{sec:construction} and the scaling of the number of parameters is analyzed in section \ref{sec:scaling}. We close by showing how existing circuits can be modified to capture a permutation and symmetry and a discussion of our results.

\begin{figure*}
    \includegraphics{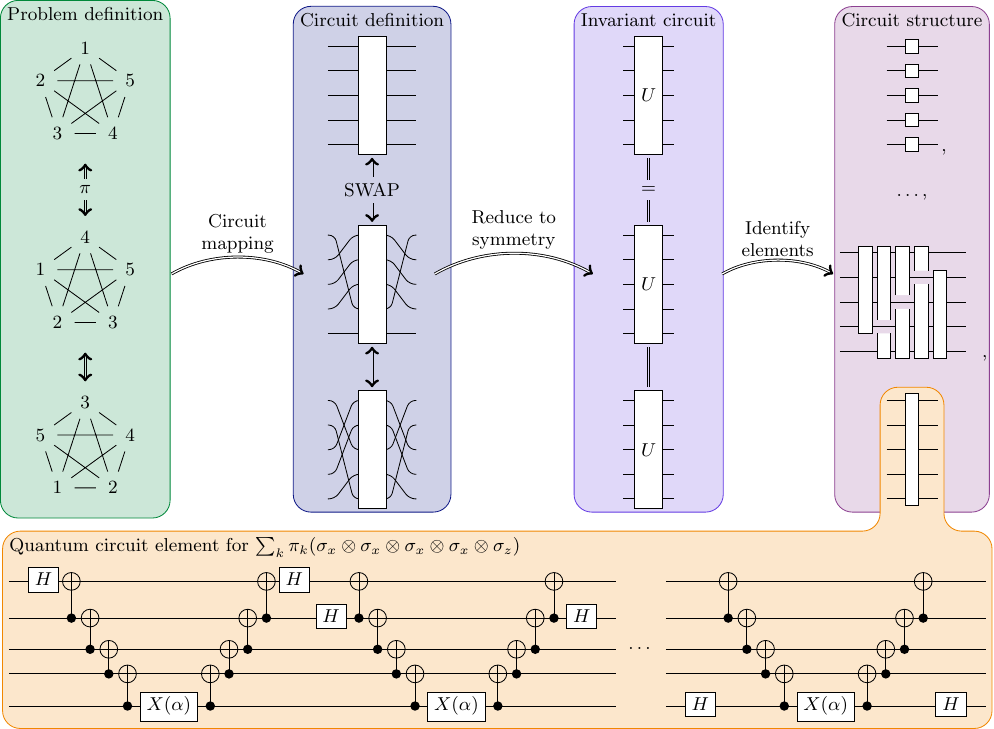}
    \caption{The main results of this paper. We show a permutation invariance in a physical problem can be implemented on an quantum computer. The permutation of inputs translates naturally to a $\SWAP$ invariance on the quantum circuit. There are circuits that are invariant under the $\SWAP$ and they form a group $piSU$. The elements of the group have a natural representation in circuit form, which in turn can be used to construct permutation invariant quantum circuits.}\label{fig:main-result}
\end{figure*}

\section{Related work}

The integration of discrete symmetries has been discussed mainly within their application to quantum machine learning by the parallel independent work of Larocca et al.~\cite{larocca_group-invariant_2022} and Meyer et al.~\cite{meyer_exploiting_2023}. They both set the basis for principally integrating discrete and continuous symmetries as restrictions on the group of quantum computing, the special unitary group $SU$. Though their work is thorough in the construction of the theory, neither of them explicitly constructs quantum circuits.

Discrete symmetries in general and permutation symmetries in general are also relevant for quantum physics. Here, it is less the computation and rather the quantum state that is required to exhibit some form of symmetry. Permutation symmetric quantum states can be used to study entanglement properties \cite{dur_three_2000, markham_entanglement_2011}. Symmetry requirements of composite systems are also used to reason about the mutual independence of systems \cite{renner_symmetry_2007}.

There is a corresponding effort in classical machine learning to integrate symmetries in the structure of the model. The most ubiquitous structure is the use of convolutional neural networks to implement a translation symmetry across a multidimensional domain \cite{wiatowski_mathematical_2018}. This type of neural network allows one to identify a signal approximately independent of its position within a larger block of data. That idea is further extended to implement other symmetries through group-invariant convolutional neural networks \cite{bronstein_geometric_2021}. The overall area of implementing symmetries into neural networks is called geometric deep learning \cite{bronstein_geometric_2021}. Within the field, graph neural networks are an extreme in this regard, designed to encapsulate the permutation symmetry inherent to many graphs \cite{zhou_graph_2020}. 


\section{Permutation-invariant quantum circuits and their associated algebras}

All operations on a quantum circuit acting on $n$-qubits correspond to an element of the \textit{special unitary} group $\operatorname{SU}(2^n)$, generated by the Lie algebra $\mathfrak{su}(2^n)$. In this section, we will provide a definition of what we mean by permutation invariant circuits at the group level. Then, by looking at the linearization of the corresponding group at the identity, i.e. its Lie algebra, the definition of permutation invariance will be extended to the relevant Lie algebra. The main result of this section will provide the connection between permutation invariant groups and permutation invariant algebras.

Before moving into any constructions or proofs, it is necessary to introduce some important definitions which will be used throughout the paper and some notation, which will greatly simplify calculations. It is also important to note that even though the choice of basis plays a role in the representation of individual objects such as matrices and vectors, physical conclusions are independent of the choice of basis. It is possible to construct the same result in any other basis with the same results. Nonetheless, to carry any concrete calculations it is necessary to pick a basis, and from now on we will adhere to the so-called \textit{Pauli Basis}, which is also the standard across quantum computing literature \cite{nielsen_quantum_2010}.

\begin{definition}[Pauli Basis]
Given an $n$-qubit system, the Pauli basis for the generators of the Lie group $\operatorname{SU}(2^n)$ consists of all $n$-tensor products of the Pauli matrices $\{\sigma_x,\sigma_y,\sigma_z, \mathbbm{1}\}$. There are in total $4^n-1$ such elements, which we will denote using the following notation for generalized Pauli matrices:

\begin{equation}	x_k y_{k-1}=\mathbbm{1}\otimes\cdots\otimes\sigma_y \otimes\sigma_x\otimes\mathbbm{1}\otimes\cdots\otimes\mathbbm{1}
\label{eq:notation}
\end{equation}
where the Pauli matrix $\sigma_x$ acts on the $k$th qubit and the $\sigma_y$ matrix acts on the $k-1$th qubit. Matrices for the rotations around $\sigma_z$ are constructed similarly and denoted by $z_k$. We will refer to a vector written in the above form as a \textit{Pauli string}.
\end{definition}

Any element in the $\mathfrak{su}(2^n)$ algebra can be written as a unique linear combination of such Pauli strings:
\begin{equation}
\mathfrak{su}(2^n) = \operatorname{span}\left\{ \bigotimes_i^n \sigma_i\right\}\backslash \mathbbm{1}^n.
\end{equation}
Any element in the $\operatorname{SU}(2^n)$ group can then be generated from this basis via the exponential map. The identity matrix $\mathbbm{1}$ is not part of the algebra since it is not a traceless matrix. 

The permutation group that we want to include as a restriction on the special unitary group $SU$ consists of all possible permutations $\pi_k$ between all available indices. On a quantum computer, the natural choice to reflect permutation is a permutation of the individual qubits carrying out the calculation. The order of qubits can be changed with a $\SWAP$ matrix.

The SWAP operation in the Pauli basis is given as the two-qubit unitary operation represented by the following matrix:
\begin{equation}
\SWAP_{1,2} =
    \begin{pmatrix}
        1 & 0 & 0 & 0\\
        0 & 0 & 1 & 0\\
        0 & 1 & 0 & 0\\
        0 & 0 & 0 & 1
    \end{pmatrix}
\end{equation}

Note that the SWAP gate is a self-inverse operation, such that $\SWAP_{i,j} \SWAP_{j,i} = \SWAP_{i,j} \SWAP_{i.j} = \mathbbm{1}$. We refer to the set of all possible $\operatorname{SWAP}$ gates in an $n$-qubit system as the $\operatorname{SWAP}$-group. This allows us to define a permutation invariance on the quantum computation.

\begin{definition}[Permutation Invariant Circuit]
A circuit $U \in SU(2^n)$ is said to be permutation invariant if it is fixed under the adjoint action of any $n$-ary swap:
\begin{align}
    \operatorname{SWAP}^{\otimes n} U \left[\operatorname{SWAP}^{\otimes n}\right]^\dagger=  U \label{eq:main-definition}
\end{align}
\end{definition}

\begin{definition}[pi$SU$]
Given an $n$-qubit system, we call the mathematical space related to a permutation invariant quantum circuit a permutation invariant special unitary group $piSU$ and its corresponding algebra $pi\mathfrak{su}$. 
\begin{equation*}
    \textnormal{pi}SU = \{ S A S=A | \forall A \in SU(2^n) \wedge \forall S\in   \SWAP \textnormal{-group} \}
\end{equation*}
\end{definition}

We suggest a pronunciation with a lax sound, such as "pin" \textipa{[pI]}, rather than a tense sound as in $\pi SU$ \textipa{[paI]}.


    It is straightforward to show that this subspace of $SU(2^n)$ is in fact a subgroup. To show this, we must show that $piSU$ is non-empty and closed under products and inverse. $piSU$ is trivially non-empty as it contains the identity element, since $\mathbbm{1}= \SWAP_{i,i}$. It remains to show closeness under product and inverse. Let $A,B$ be arbitrary elements of $piSU$, using the definition of $piSU$ and the self-inverse property of $\SWAP$ operators, we show that $piSU$ is closed under group multiplication and inverse element:
    
\begin{align*}
    & \SWAP_{i,j} \left( AB\right) \SWAP_{i,j}\\
     &=\SWAP_{i,j} \left( \SWAP_{i,j} A \SWAP_{i,j} \right)B \SWAP_{i,j}\\
     & = A \left (\SWAP_{i,j} B \SWAP_{i,j} \right)= AB,
\end{align*}
and

\begin{align*}
    & \mathbbm{1}= \SWAP_{i,j} \left( A A^{-1}\right) \SWAP_{i,j}\\
    & =  \SWAP_{i,j} \left( \SWAP_{i,j} A \SWAP_{i,j} \right) A^{-1}\SWAP_{i,j}\\
    & = A \left( \SWAP_{i,j} A^{-1} \SWAP_{i,j} \right)\\
    & \Rightarrow A^{-1}= \SWAP_{i,j} A^{-1} \SWAP_{i,j}.
\end{align*}

 Thus, we have shown that $piSU$ is a closed subgroup of $SU(2^n)$. Moreover, Cartan's closed-subgroup theorem states that if $H$ is a closed subgroup of a Lie group $G$, then $H$ is an embedded Lie subgroup, see chapter 5 in \cite{lee_introduction_2012} for a detailed proof. Moreover, the subgroup-subalgebra correspondence theorem states that there is a unique Lie sub-algebra generating the subgroup via the exponential map \cite{lee_introduction_2012}.

Since it has been shown that $piSU$ is a closed subgroup of $SU(2^n)$, there must be a unique Lie sub-algebra of $\mathfrak{su}$, which we will refer to as $pi\mathfrak{su}$, that generates $piSU$.

\begin{figure}[h]
    \centering
    \includegraphics{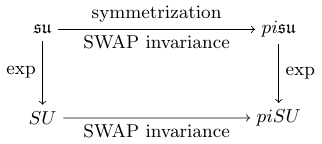}
    \caption{Diagram of the relation between the algebra $\mathfrak{su}$ and its group $SU$. One can move from the general $SU$ group to the permutation invariant group $piSU$ with the SWAP invariance and from the algebra $\mathfrak{su}$ to the algebra $pi\mathfrak{su}$ with a complete symmetrization. The exponential map connects the algebra $pi\mathfrak{su}$ to the group $piSU$ exactly.}
    \label{fig:algebragroupcon}
\end{figure}

Using the definition of the Lie algebra of a Lie group as the linearization of the group around the identity $\mathbbm{1}$, we will proceed to explicitly construct the $pi\mathfrak{su}$ sub-algebra. Given the smooth structure of $piSU$ as an embedded subgroup, pick a smooth path $\gamma(t)$ going through the origin so that $\gamma(0)=\mathbbm{1}$ and $\gamma(t) \in piSU$ for all $t \in \mathbbm{R}$. To first order, one can expand around the origin as follows:
\begin{equation}
    \gamma(t)= \gamma(0)+ t \frac{d A(t)}{dt}|_{t=0}+ \mathcal{O}(t^2)
\end{equation}
By definition $\frac{d A(t)}{dt}|_{t=0} \in pi\mathfrak{su}$. Let us denote $\frac{d A(t)}{dt}|_{t=0}=B$, combining it with the above equation gives:

\begin{equation}
    \gamma(t)= \mathbbm{1}+tB+\mathcal{O}(t^2)
\end{equation}
Imposing the condition of invariance under the adjoint action of any element $S$ of the $\operatorname{SWAP}$-group gives the following constraint of the elements of the Lie algebra:
\begin{equation}
    B=SBS, \  \ \forall B \in pi\mathfrak{su}.
\end{equation}
Therefore, the elements of the $pi\mathfrak{su}$ sub-algebra are skew Hermitian and traceless matrices invariant under the adjoint action of the $\operatorname{SWAP}$-group. The relations among the discussed algebras and groups are depicted in figure \ref{fig:algebragroupcon}.

First, we note that given an arbitrary basis vector of $pi\mathfrak{su}$ written in the form shown in equation \ref{eq:notation},  the adjoint action of a $\operatorname{SWAP}$-operator results in a change of order in the Pauli string, for instance:
\begin{equation}
    \operatorname{SWAP}_{i,j} (x_i y_j) \operatorname{SWAP}_{i,j}= x_jy_i.
\end{equation}

Thus, if we construct a new basis, whose elements are defined as the symmetric sum of all Pauli strings containing a fixed number of $\sigma_x,\sigma_y,\sigma_z, \mathbbm{1}$, such a basis will by construction be $\operatorname{SWAP}$-invariant. Consisting of fully symmetric linear combinations of $\mathfrak{su}$ vectors, the resulting space meets all the desired properties for $pi\mathfrak{su}$; skew Hermitian, traceless and $\operatorname{SWAP}$-invariant. Finally, it remains to show that the constructed space is in fact a Lie subalgebra, i.e. that is it is closed under the Lie bracket. Below, we provide an explicit proof for $pi\mathfrak{su}(4)$. The concept of this proof can be extended to higher $n$, however, it becomes increasingly difficult to keep track of the indices, so we conjecture that for a given $n$, it is possible to exhaustively compute all the brackets explicitly, but we provide no general proof for arbitrary $n$.\\

\begin{proof2}[The algebra of $pi\mathfrak{su}(4)$ is closed.]

First of all, notice that all basis elements of $pi\mathfrak{su}(4)$ can be written, up to constants, either as $(\sigma_a \otimes \sigma_b + \sigma_b \otimes \sigma_a) $ or $(\sigma_a \otimes \mathbbm{1} + \mathbbm{1} \otimes \sigma_a)$. Thus, to show that $pi \mathfrak{su}(4)$ is a subalgebra of $\mathfrak{su}(4)$ one needs to show that the following commutators are closed under the Lie bracket:
\begin{align}
    & [(\sigma_a \otimes \sigma_c + \sigma_c \otimes \sigma_a), (\sigma_b \otimes \sigma_d + \sigma_b \otimes \sigma_d)] \\ 
    \label{eq:commutator2}
    & [(\sigma_a \otimes \mathbbm{1} + \mathbbm{1} \otimes \sigma_a), (\sigma_b \otimes \mathbbm{1} + \mathbbm{1} \otimes \sigma_b)]\\
    \label{eq:commutator3}
    &[(\sigma_a \otimes \sigma_c + \sigma_c \otimes \sigma_a),(\sigma_b \otimes \mathbbm{1} + \mathbbm{1} \otimes \sigma_b) ].
\end{align}
Using the structure constants $\epsilon$ of the Lie algebra 
$\mathfrak{su}(2)$ and the anti-symmetry of the commutator, it is possible to simplify the following commutators:

\begin{equation*}
[\sigma_i \otimes \sigma_k, \sigma_j \otimes \sigma_l ]= 2 i \epsilon_{ija}\delta_{kl}(\sigma_a \otimes \mathbbm{1})+ 2i \epsilon_{klb}\delta_{ij}(\mathbbm{1} \otimes \sigma_b).
\end{equation*}

Using the above result, we find the following expressions for the commutators:
\begin{align*}
     [(\sigma_a \otimes \sigma_c + \sigma_c \otimes \sigma_a), &(\sigma_b \otimes \sigma_d + \sigma_d \otimes \sigma_b)]\\
    &=2 i \epsilon_{abe} \delta_{cd}(\sigma_e \otimes \mathbbm{1} + \mathbbm{1} \otimes \sigma_e) \\
    &+2 i \epsilon_{cde} \delta_{ab}(\sigma_e \otimes \mathbbm{1} + \mathbbm{1} \otimes \sigma_e) \\
    &+2 i \epsilon_{ade} \delta_{cb}(\sigma_e \otimes \mathbbm{1} + \mathbbm{1} \otimes \sigma_e) \\
    &+2 i \epsilon_{cbe} \delta_{ad}(\sigma_e \otimes \mathbbm{1} 
    + \mathbbm{1} \otimes \sigma_e)
\end{align*}
\begin{align*}
    [(\sigma_a \otimes \mathbbm{1} + \mathbbm{1} \otimes \sigma_a), &(\sigma_b \otimes \mathbbm{1} + \mathbbm{1} \otimes \sigma_b)]\\
    & = 2i \epsilon_{abc}(\sigma_c \otimes \mathbbm{1} + \mathbbm{1} \otimes \sigma_c)
\end{align*}
\begin{align*}
[(\sigma_a \otimes \sigma_c + \sigma_c \otimes \sigma_a),&(\sigma_b \otimes \mathbbm{1}
+ \mathbbm{1} \otimes \sigma_b) ]\\
&=2 i \epsilon_{abd}(\sigma_d \otimes \sigma_c + \sigma_c \otimes \sigma_d)\\
&+2 i \epsilon_{cbd}(\sigma_d \otimes \sigma_a + \sigma_a \otimes \sigma_d)
\end{align*}

It is straightforward to see that for all $a,b,c,d \in \{x,y,z\}$ the above commutators are elements of $pi\mathfrak{su}(4)$.  Thus, we have proved that $pi\mathfrak{su}(4)$ is closed under the Lie bracket of $\mathfrak{su}(4)$, and thus, a subalgebra of $\mathfrak{su}(4)$.
\hfill $\square$
\end{proof2}

These results demonstrate that the $pi\mathfrak{su}$ Lie subalgebra is spanned by symmetrized Pauli strings:
\begin{equation}
pi\mathfrak{su}(2^n) = \operatorname{span}\left\{ \sum_k \bigotimes_i^n\pi_k\sigma_i\right\}\backslash \mathbbm{1}^n.
\end{equation}

The permutation $\pi_k$ and the summation $\sum_k$ iterate over all possible permutations of $[x, y, z, \mathbbm{1}]$ in the Pauli string. Each element in $pi\mathfrak{su}$ corresponds to the sum of all permutations of the Pauli matrices contained within its Pauli string. We show this construction explicitly for the two-qubit case $piSU(4)$ in section \ref{ssec:pisu4}.


The definition of a SWAP-invariant group can be generalized to any other invariance based on swapping entries of the matrix. The CNOT matrices,
\begin{align*}
    \CNOT_{1,2} = \begin{pmatrix}
        1 & 0 & 0 & 0\\
        0 & 0 & 0 & 1\\
        0 & 0 & 1 & 0\\
        0 & 1 & 0 & 0
    \end{pmatrix},\quad
        \CNOT_{2,1} = \begin{pmatrix}
        1 & 0 & 0 & 0\\
        0 & 1 & 0 & 0\\
        0 & 0 & 0 & 1\\
        0 & 0 & 1 & 0
    \end{pmatrix}
\end{align*}
individually act as swapping matrices on the index 2 and 4, 3 and 4 respectively. It is possible to then construct a group that is invariant under each operation – $\CNOT_{1,2}$- and $\CNOT_{2,1}$-invariant subgroups of $SU(4)$ and their higher order analogues. Similarly, one can imagine permutation matrices that act on any two indices. Effectively these invariant groups are all versions of the quotient group $SU(2^n)/S_n$, the special unitary group of size $2^n$ with $n$ arbitrarily interchangeable indices. The corresponding quantum circuits may not always be easily constructable nor have a direct application. They do however follow the same parameter scaling structure of section \ref{sec:scaling} and have the same overall mathematical structure.

\section{Construction of circuits with permutation symmetry}\label{sec:construction}

The elements of $piSU$ can be expressed as quantum circuits, as any element of $SU$ can \cite{mansky_near-optimal_2023}. Using the algebra $pi\mathfrak{su}$, it is possible to construct most of the elements directly and build circuits schematically that are SWAP invariant. There is a direct construction of a quantum circuit element from a given element of the algebra \cite{mansky_decomposition_2023}.

\subsection{Full classification for $piSU(4)$}\label{ssec:pisu4}

As an example, we provide a complete classification of $piSU(4)$, the permutation invariant group for two qubits. The structure of the $piSU$ group can be understood through its accompanying Lie algebra, $pi\mathfrak{su}(4)$. Its basis are the symmetrized elements of the overalgebra:
\begin{align}
    pi\mathfrak{su}(4) = \sum_k \bigotimes^2_{i=1} \pi_k(\sigma_i)
\end{align}

This means that each combination of Pauli matrices is symmetrized:

\begin{align*}
    \sigma_1 &= \sigma_x \otimes \sigma_x\\
    \sigma_2 &= \sigma_x \otimes \sigma_y + \sigma_y \otimes \sigma_x\\
    \sigma_3 &= \sigma_x \otimes \sigma_z + \sigma_z \otimes \sigma_x\\
    \sigma_4 &= \sigma_x \otimes \mathbbm{1} + \mathbbm{1} \otimes \sigma_x\\
    \sigma_5 &= \sigma_y \otimes \sigma_y\\
    \sigma_6 &= \sigma_y \otimes \sigma_z + \sigma_z \otimes \sigma_y\\
    \sigma_7 &= \sigma_y \otimes \mathbbm{1} + \mathbbm{1} \otimes \sigma_y\\
    \sigma_8 &= \sigma_z \otimes \sigma_z\\
    \sigma_9 &= \sigma_z \otimes \mathbbm{1} + \mathbbm{1} \otimes \sigma_z
\end{align*}

For $pi\mathfrak{su}(4)$, it is possible to construct all the circuits directly from the algebra as there are only two Pauli matrices in each string. These are shown in table \ref{tab:su4-drawings}. Each string has a central $\sigma_x \otimes \sigma_x$ or $\sigma_z \otimes \sigma_z$ structure as its backbone. The axis of each Pauli matrix can be changed with $H$ and $S, S^\dagger$ gates \cite{nielsen_quantum_2010}.  These circuits can be individually and combination be verified to fulfil equation \eqref{eq:main-definition}. Their construction can be explained intuitively as a central entangling structure generated by $\bigotimes \sigma_x$ or $\bigotimes \sigma_z$ with a change of basis for the corresponding qubit.

\begin{table*}
    \centering
        \caption{Corresponding circuits for the exponentials of the algebra elements in $pi\mathfrak{su}(4)$ as $\exp(-i\alpha/2 \sum_k \pi_k\bigotimes_{i=1}^2 \sigma_i)$. The circuits are constructed on a $\sigma_x \otimes \sigma_x$ backbone with changes to the qubit axis as necessary. For some circuits, there are optimized visions with fewer gates available.}
\begin{tabularx}{13.5cm}{r l l}
Exponent & Circuit & Optimized circuit\\\hline
$\sigma_x \otimes \sigma_x$ &\includegraphics{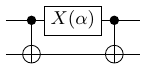}\\
$\sigma_x \otimes \sigma_y + \sigma_y \otimes \sigma_x$ & \includegraphics{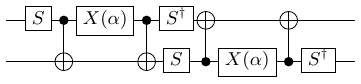}\\
$\sigma_x \otimes \sigma_z + \sigma_z \otimes \sigma_x$ & \includegraphics{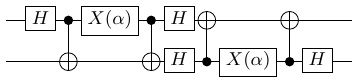} & \includegraphics{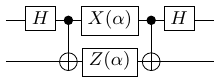}\\
$\sigma_x \otimes \mathbbm{1} + \mathbbm{1} \otimes \sigma_x$ & \includegraphics{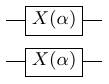}\\
$\sigma_y \otimes \sigma_y$ & \includegraphics{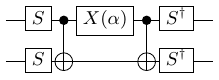}\\
$\sigma_y \otimes \sigma_z + \sigma_z \otimes \sigma_y$ & \includegraphics{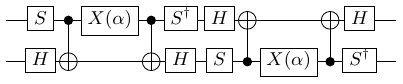}\\
$\sigma_y \otimes \mathbbm{1} + \mathbbm{1} \otimes \sigma_y$ & \includegraphics{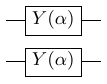}\\
$\sigma_z \otimes \sigma_z$ & \includegraphics{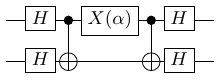} & \includegraphics{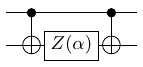}\\
$\sigma_z \otimes \mathbbm{1} + \mathbbm{1} \otimes \sigma_z$ & \includegraphics{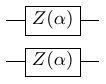}\\\hline
\end{tabularx}

    \label{tab:su4-drawings}
\end{table*}

During exponentation, each term of the permutation-invariant algebra has a complex-valued parameter $\alpha$ that corresponds to the parameter generating the $X$ rotation in the group. The parameterized gates allow the full $piSU$ group coverage.

The individual circuits can also be optimized to reduce the number of basic operations within them. It is known that any two-qubit operation can be expressed with at most three qubits \cite{vatan_optimal_2004, shende_minimal_2004}. The four-qubit circuits of table \ref{tab:su4-drawings} are not optimal in this sense. For symmetry reasons, the circuits should be expressible with two $\CNOT$s only.

\subsection{General construction of elements}\label{ssec:general-construction}

\begin{figure}
    \centering
\includegraphics{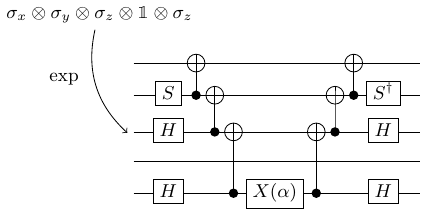}
    \caption{Example of a Pauli string exponential represented as a quantum circuit. On the backbone of a central $\bigotimes_{i=1}^5 \sigma_x$ structure that creates the central $\CNOT$ ladder, each qubit gets changed to its appropriate $\sigma_y$, $\sigma_z$ by a $S, S^\dagger$, $H, H$ pairs on the outside of the ladder, respectively. Identity elements $\mathbbm{1}$ in the Pauli string are created by skipping the respective qubit.}
    \label{fig:zzyz}
\end{figure}

The construction of the quantum circuit corresponding to an individual Pauli string in the algebra is straightforward. The group element is expressed as $\exp(-i\alpha/2\bigotimes_i^n \sigma_i)$. The parameter $\alpha$ is expressed through a rotation gate on the $X$ axis. The connection to the other qubits is established through CNOT ladders that flow out on both sides of the rotation gate. The axis for each Pauli gate corresponding to a qubit can be changed with $H$ for $Z$ axis and $S, S^\dagger$ for the $Y$ axis. Qubits with a $\mathbbm{1}$ in their corresponding Pauli matrix are not addressed. An example of this structure is shown in figure \ref{fig:zzyz}.

The arrangement of $\CNOT$s in the interior is not fixed to a broadening ladder from one qubit to another. All rearrangements of the qubits are equivalent, a direct result of the qubit permutation invariance. The ladder can also be restructured to connect to the rotation matrix directly or form other shapes, as long as the $\CNOT$ layers are connected to the central rotation gate increasing from the inside \cite{mansky_decomposition_2023, cowtan_phase_2020}.

For a sum of Pauli strings, this approach is not directly extensible. An exponential of a sum is not trivially the same as a product of exponentials. However, results from quantum error correction, in particular stabilizer codes \cite{gottesman_stabilizer_1997} show that if the individual elements within the sum in the algebra commute, the individual Pauli strings can be exponentiated and combined in the group. The Pauli strings commute if only two of three Pauli matrices $\{\sigma_x, \sigma_y, \sigma_z\}$ are present \cite{djordjevic_quantum_2022}.

\begin{figure}
    \centering
\includegraphics{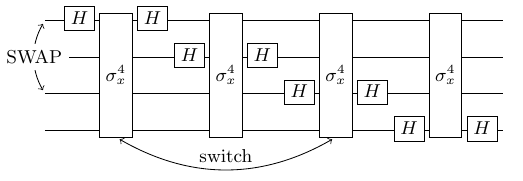}

    \caption{The $\SWAP$ operations between qubits are equivalent to changes in order between the individual elements of the exponential sum of Pauli string. The underlying $\bigotimes \sigma_x$ element between the pairs of Hadamard gates $H$ is intrinsically invariant under the $\SWAP$ operation.}
    \label{fig:swaps}
\end{figure}

The result that exponents with only two Pauli matrices in the strings commute also has an intuitive explanation: The vertical $\operatorname{SWAP}$ operations between the qubits are equivalent to horizontal swaps between the individual elements. This is illustrated in figure \ref{fig:swaps}. This property implies that the individual elements of the product of exponentials commute. In the case of Pauli strings consisting of all components, the number of strings is greater than the number of available $\SWAP$s and it is no longer possible to arrange the individual strings in an arbitrary order.

In the case where the Pauli strings are directly expressible through exponentiation, it possible to optimize the resulting CNOT ladders for shorter circuit depths \cite{cowtan_phase_2020}. The authors achieve a 50\% reduction in the number of CNOTs employed in the test circuits that they use. The naive structure for circuits requires $2n\sum_{k=1}^n (n-k)^3 = \mathcal{O}(n^5)$ CNOTs for construction of the whole group.

The elements of $piSU$ always affect all qubits, but their connectivity varies depending on how many qubits are addressed at the same time. The Pauli strings containing only Pauli matrix and identity elements $\mathbbm{1}$ otherwise have the lowest connectivity, with their corresponding circuits consisting of equivalent rotation matrices on all qubits only. On the other end, a Pauli string of only Pauli matrices links all qubits together with $\CNOT$ gates. In between, for $k$ Pauli matrices present in the individual Pauli string, $k$ out of $n$ qubits are addressed. This is termed locality and restrictions of locality can be important on some quantum computers. While restrictions of locality almost have no effect on the expressibility of the circuit \cite{divincenzo_two-bit_1995, lloyd_almost_1995}, the same is not true for permutation symmetric quantum gates with locality restrictions \cite{marvian_restrictions_2022,kazi_universality_2023}. Here it is necessary to have all elements of the algebra present in the group to achieve full expressibility. 

From a circuit diagram, it is not visually obvious whether a given quantum circuit is $\SWAP$ invariant or not. This is a general drawback of circuit diagrams, in the standard language used in this work as well as its varieties such as ZX-calculus \cite{van_de_wetering_zx-calculus_2020}. Within ZX-calculus, operators that modify all qubit lines equally and are therefore permutation invariant have been proposed to be represented by their own special symbol \cite{cowtan_phase_2020}. In the standard notation, no such symbol is available.

\section{Scaling behaviour}\label{sec:scaling}


The group $piSU(2^n)$ captures  how unitary transformations on $n$-qubits behave under permutation symmetries and it provides insights into the inherent structure and relationships between unitary operators in the presence of permutation operations. In the following lemma, we establish a fundamental result that characterizes the dimension of the group $piU(2^n)$, where we show that the dimension of $piU(2^n)$ is given by $(n+3)(n+2)(n+1)/6$.

The dimension of the two-qubit unitary Lie group $U(4)$ can be determined by counting the Pauli matrices spanning its Lie algebra, $\sum_{i,j=0}^{3}\sigma_i\otimes\sigma_j$, giving dim $U(4)=4^2$. In order to determine the dimension of the two-qubit permutation-invariant group $piU(4)$ we need to account for the symmetry present and account for the degenerate elements.

The group $piSU(4)$ then has 9 dimensions, due to its unitarity constraint. The number of elements generalizes via the multinomial theorem \cite{brualdi_introductory_2008} to
\begin{align}
    pi(\sigma_i)^{\otimes n} = \sum_{\substack{i_1,i_2,i_3,i_4\geq 0 \\ i_1+i_2+i_3+i_4= n}}\binom{n}{i_1,i_2,i_3,i_4}\sigma_x^{i_1}\sigma_y^{i_2}\sigma_z^{i_3}1^{i_4}
\end{align}

The number of terms in such a multinomial sum is equal to the number of monomials of $n$th-degree on the variables $x,y,z,1$.

Since we are determining the dimension of the algebra spanned by the independent terms, each degenerate term needs to be counted as one,
\begin{equation}
    \#piSU(2^n)=\binom{n+4-1}{4-1} - 1=\frac{(n+3)(n+2)(n+1)}{6} - 1
\end{equation}

The number of independent terms grows as $n^3$ for the permutation invariant special unitary group. 
\begin{align}
    \dim(piSU(2^n)) = \mathcal{O}(n^3)
\end{align}

This is significantly less than the number of parameters in the full, non-symmetric group $\dim(SU(2^n)) = \mathcal{O}(4^n)$. It is a consequence of the fact that calculations executed on a quantum computer that exhibit this symmetry group cannot make use of the non-symmetric part of the Hilbert space. These circuits are constrained to the symmetric part and their scaling behaviour with respect to the number of qubits is severely limited. This is true for both quantum machine learning approaches utilizing an ansatz as well as schematic approaches that prescribe a construction of gates on the qubits. It has been shown that the restrictiveness of the symmetry prevents a quantum speed-up relative to classical computers \cite{ben-david_symmetries_2020}.






\section{Circuit modification}\label{sec:applications}

\begin{figure}
\centering
\includegraphics{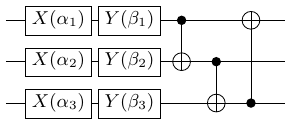}

    \caption{An example of a variational circuit of three qubits that is subsequently modified to be permutation invariant with a copy of itself and intrinsically permutation invariant between all qubits. The circuit is constructed of two layers of rotation matrices with parametrized angles and a layer of $\CNOT$ elements that provide entanglement. Several of these elements can be concatenated to form a variational quantum circuit \cite{schuld_introduction_2015}.}
    \label{fig:var-circuit-base}
\end{figure}

In this section we show how the permutation symmetry can be implemented into a quantum circuit ansatz. On the basis of a commonly used variational circuit element, we show how to extend a circuit to have permutable blocks of information and how to modify an existing circuit to capture permutation symmetry across all circuits. Adopting a known circuit to exhibit permutation symmetry allows us to capture the present knowledge about the behaviour of the quantum circuit and extend it to recognize the underlying symmetry. The starting circuit in both cases is shown in figure \ref{fig:var-circuit-base}. It consists of two layers of parametrized rotation matrices and a strongly entangling layer with $\CNOT$s \cite{schuld_introduction_2015}.

It is also possible to construct a quantum circuit ansatz by combining the elements that are presented above. Due to the group properties of $piSU$, the combination of elements in any order is still a permutation symmetric circuit.

\subsection{Symmetry by extending a circuit}

\begin{figure}[h]
    \centering
    \includegraphics{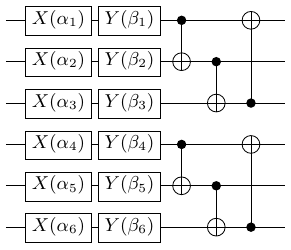}
    \caption{The starting point for a permutation symmetric quantum circuit by extension. The base circuit of figure \ref{fig:var-circuit-base} is doubled up into two separate quantum circuits. The two blocks need to be connected to be permutation invariant.}
    \label{fig:variational-circuit}
\end{figure}

In many quantum algorithms, the mapping of information to qubits is not one-to-one. Rather, input is expressed across several qubits, but the input itself may exhibit a permutation symmetry.  We illustrate the procedure by considering an extension of the starting variational circuit of figure \ref{fig:var-circuit-base}. The circuit is modified by doubling the circuit to mimic two permutation invariant blocks. Now each of the two blocks of three qubits corresponds to some information, but the two blocks are interchangeable. An illustration of that starting circuit is shown in figure \ref{fig:variational-circuit}. The symmetrization requirement is then $\{\SWAP_{1,4}, \SWAP_{2,5}, \SWAP_{3, 6}\}$.

\begin{figure}[h]
    \centering
\includegraphics{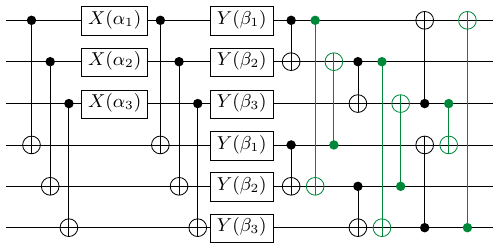}
    \caption{The modified variational circuit for two permutation symmetric blocks of information. The parametrized $X$ and $Y$ rotations can be made permutation invariany by coupling them to $\sigma_x \otimes \sigma_x$, shown for the $X$ rotation, or by equalizing their rotation parameters, corresponding to $\sigma_y \otimes \mathbbm{1} + \mathbbm{1} \otimes \sigma_y$ for the $Y$ rotation. The entangling layer of $\CNOT$s incorporates the additional coonections that would appear under a $\SWAP$. These additional $\CNOT$ gates are drawn in green.}
    \label{fig:gate-difference}
\end{figure}

The first rotations in the $X$ and $Y$ axis can be symmetrized either as $\sigma_x \otimes \mathbbm{1} + \mathbbm{1} \otimes \sigma_x$ or as $\sigma_x \otimes \sigma_x$ and $\sigma_y \otimes \mathbbm{1} + \mathbbm{1} \otimes \sigma_y$ or $\sigma_y \otimes \sigma_y$ respectively. The first option corresponds to setting the corresponding parameters in each block to the same value, such that $\alpha_i = \beta_i$, the second one introduces additional entanglement between the two blocks via $XX$ ($YY$) gates. The difference is illustrated in figure \ref{fig:gate-difference}, where the $X$ gates are symmetrized as $\sigma_x \otimes \sigma_x$ and the $Y$ gates as $\sigma_y \otimes \mathbbm{1} + \mathbbm{1} \otimes \sigma_y$. The two approaches are not equivalent. Depending on the application, one or the other may be preferable.

The second part of the variational circuit, the dense entanglement layer, also requires symmetrization. In contrast to the strongly entangling layer itself above, two layers or more layers can be made permutation invariant. Similar to the horizontal and vertical permutation of elements shown in section \ref{ssec:general-construction}, a vertical permutation of elements through $\SWAP$ operations implies a horizontal exchange of elements. The circuit needs to be expanded to also contain the elements that would be generated under an exchange of qubits. All $\CNOT$ gates that would appear under a $\SWAP$ operation are present in the circuit. The green $\CNOT$ gates added to the pairs of $\CNOT$ gates are abelian with the original ones, therefore the order of the $\CNOT$ gates does not matter. 

The final circuit in figure \ref{fig:gate-difference} is invariant under the $\SWAP$ group $\{\SWAP_{1, 4}, \SWAP_{2, 5}, \SWAP_{3, 6}\}$. It is not invariant under $\SWAP$ operations between qubits within a block. The permutation-invariant circuits have the same number of parameters as the starting circuit shown in figure \ref{fig:var-circuit-base}, independent of the number of additional blocks added to the overall circuit. This a consequence of the restricted choices for parameterized gates in the direct translation. More parameters can be added to circuit by concatenating some of the permutation invariant quantum circuit elements presented above.

\subsection{Symmetry by changing a circuit}

In the case where one has an existing circuit and now wants to impose a permutation symmetry on all inputs and outputs of the circuits, the circuit ansatz needs to be symmetrized. Rather than constructing a new circuit from scratch, the knowledge obtained so far can be used in the modified circuit. We illustrate the methods on a simple yet common variational quantum circuit with three qubits, consisting of two layers of rotation matrices and a dense entanglement layer \cite{schuld_variational_2021}. The base circuit is illustrated in figure \ref{fig:var-circuit-base}. These first two layers of single-qubit rotations contain the parameters of the circuit that are optimized during a training loop. The symmetrization requirement is that all qubits can be permuted, such that the circuit is invariant under any combination of $\{\SWAP_{1,2}, \SWAP_{2,3}, \SWAP_{1,3}\}$.

\begin{figure}
    \centering
\includegraphics{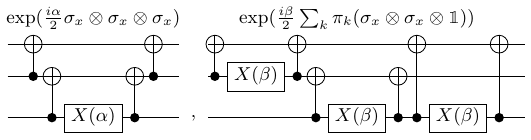}
    \caption{The circuits corresponding to $\sigma_x \otimes \sigma_x \otimes \sigma_x$ and $\sigma_x \otimes \sigma_x \otimes \mathbbm{1} + \sigma_x \otimes \mathbbm{1} \otimes \sigma_x + \mathbbm{1} \otimes \sigma_x \otimes \sigma_x$. }
    \label{fig:xx1-xxx}
\end{figure}

The circuit can be adapted to reflect a permutation symmetry by working layer by layer. The one-qubit rotations express the algebra element $\alpha_1\sigma_{x} \otimes \mathbbm{1} \otimes \mathbbm{1} + \alpha_2 \mathbbm{1} \otimes \sigma_{x} \otimes \mathbbm{1} + \alpha_3 \mathbbm{1} \otimes \mathbbm{1} \otimes \sigma_{x}$. The obvious approach to symmetrization is then to set the values of the parameters $\alpha_1 = \alpha_2 = \alpha_3$. This corresponds directly to the structure present in the original ansatz. An alternative is to implement $\alpha \sigma_x \otimes \sigma_x \otimes \sigma_x$ or $\alpha (\sigma_x \otimes \sigma_x \otimes \mathbbm{1} + \sigma_x \otimes \mathbbm{1} \otimes \sigma_x + \mathbbm{1} \otimes \sigma_x \otimes \sigma_x$. These structures provide a different path through the available Hilbert space and can be more suitable to some problems. These alternatives are illustrated in figure \ref{fig:xx1-xxx}.

For the second part of the starting variational circuit of figure \ref{fig:var-circuit-base}, the same consideration of symmetrizing across all qubits applies. This $\CNOT$ structure, often referred to as a "strongly entangling layer" \cite{bergholm_pennylane_2022}, acts to interchange superpositions between the different qubits. The construction can also be understood as a permutation matrix in $SU(2^n)/S_{2^n}$, with order $2^n-1$. This structure is not part of $piSU$ and cannot readily be made to fit into it. This is in contrast with the previous example of symmetrizing by extending a circuit. If the circuit layer is modified in the same way as above, by adding the $\CNOT$ elements that would appear under a permutation, the resulting structure is not abelian and therefore not permutation invariant.

The element can then be replaced with one from $piSU$, for example one that introduces a rotation around the central permutation invariant axis, such as $\sum_k\pi_k(\sigma_x \otimes \sigma_x \otimes \mathbbm{1})$. In cases where the number of $\CNOT$s is restricted or important, a good replacement would be $\sigma_x \otimes \sigma_x \otimes \sigma_x$. Both circuit element alternatives add entangling structures to the circuit. They also provide additional training parameters to the circuit, which were reduced in the prior rotation gate block.

The presented approach can be readily adopted for higher numbers of qubits. With increasing qubit number comes the consideration of how many $k$-local structures are necessary and reasonable to implement on the system of choice. Compared to the non-symmetrized ansatz, the fraction of parameters to non-parametrized two-qubit gates is low, but the size of the available Hilbert space is also only growing moderately.


\section{Discussion and future work}

In this work, we show the direct representation of permutational symmetries in quantum circuits. Implementing a physical symmetry in a computation is important because it allows to reduce the number of parameters in the quantum circuit while maintaining the necessary expressibility of the circuit. A general circuit that is able to cover the whole Hilbert space can achieve the same solution, but by restricting to the necessary space the quantum circuit is able to reflect the same calculation with less computational depth. 

The mathematical basis, in particular closure of the permutation invariant subgroup of $SU$, enables the construction of quantum circuits. They follow a clear structure and are generally directly expressible as quantum circuit elements. As a result, they can be directly implemented into the quantum circuit construction toolkits used for quantum machine learning and related ansatz-based approaches. 

The approach of constructing permutation invariant blocks of qubits also facilitates higher order structures on the physical qubits. The presented approach can be extended to structures of logical qubits or to incorporate quantum error correction structures. It shows that including discrete group symmetries into quantum systems is not only theoretically possible but concrete quantum circuits can be realized, at least for the permutation symmetry. 

The present work deals with permutation symmetries, which act as the supergroup for any other discrete symmetry group. This choice is deliberate, as it makes the mathematics of invariants easy to understand. However, it also limits the scope of applicability to select problems. An exploration of how to implement any discrete symmetry as a restriction on the special unitary group is possible expansion of our work here.

\begin{acknowledgements}
    The authors acknowledge funding from the German Federal Ministry of Education and Research (BMBF) under the funding program "Förderprogramm Quantentechnologien – von den Grundlagen zum Markt" (funding program quantum technologies – from basic research to market), project BAIQO, 13N16089.
\end{acknowledgements}

\bibliography{references, local-refs}

\appendix

\end{document}